\documentclass[prd,superscriptaddress,twocolumn,nofootinbib,showpacs,preprintnumbers,floatfix]{revtex4}

\usepackage{mathrsfs}
\usepackage{amsfonts,amssymb,amsmath}
\usepackage{graphicx,epsfig}
\usepackage{multirow}
\usepackage{verbatim}

\def\fsu5{$\cal{F}$-$SU(5)$}
\def\bfsu5{$\boldsymbol{\mathcal{F}}$-$\boldsymbol{SU(5)}$}

\def\m1half{$M_{1/2}$}
\def\m3half{$M_{3/2}$}
\def\m32{$M_{32}$}

\def\fb{${\rm fb}^{-1}$~}
\def\fbns{${\rm fb}^{-1}$}

\begin{document}

\title{Chanel $\mathbf{N^o5}$ ($\mathbf{{fb}^{-1}}$):\\The Sweet Fragrance of SUSY}

\author{Tianjun Li}

\affiliation{State Key Laboratory of Theoretical Physics and Kavli Institute for Theoretical Physics China (KITPC),
Institute of Theoretical Physics, Chinese Academy of Sciences, Beijing 100190, P. R. China}

\affiliation{George P. and Cynthia W. Mitchell Institute for Fundamental Physics and Astronomy,
Texas A$\&$M University, College Station, TX 77843, USA}

\author{James A. Maxin}

\affiliation{George P. and Cynthia W. Mitchell Institute for Fundamental Physics and Astronomy,
Texas A$\&$M University, College Station, TX 77843, USA}

\author{Dimitri V. Nanopoulos}

\affiliation{George P. and Cynthia W. Mitchell Institute for Fundamental Physics and Astronomy,
Texas A$\&$M University, College Station, TX 77843, USA}

\affiliation{Astroparticle Physics Group, Houston Advanced Research Center (HARC),
Mitchell Campus, Woodlands, TX 77381, USA}

\affiliation{Academy of Athens, Division of Natural Sciences,
28 Panepistimiou Avenue, Athens 10679, Greece}

\author{Joel W. Walker}

\affiliation{Department of Physics, Sam Houston State University,
Huntsville, TX 77341, USA}


\begin{abstract}
We present compounding evidence of supersymmetry (SUSY) production at the LHC, in the
form of correlations between the nominal 5 \fb ATLAS and CMS results for the $\sqrt{s} = 7$~TeV 2011 run and detailed
Monte Carlo collider-detector simulation of a concrete supersymmetric model named No-Scale \fsu5. 
Restricting analysis to those event selections which yield a signal significance $S/\sqrt{B+1}$ greater
than 2, we find by application of the $\chi^2$ statistic that strong correlations exist among
the individual search strategies and also between the current best fit to the SUSY mass scale
and that achieved using historical 1 \fb data sets.  Coupled with an appropriately large
increase in the ``depth'' of the $\chi^2$ well with increasing luminosity, we suggest that these
features indicate the presence of a non-random structure to the data -- a light
fragrance perhaps evocative of some fuller coming fruition.  Those searches having signal
significances below 2 are assembled into a lower exclusion bound on the gaugino mass, which is
shown to be consistent with the prior best fit.  Assuming the forthcoming delivery of an additional tranche
of integrated luminosity at 8 TeV during 2012 that measures on the order 15 \fbns, we project a sufficiency of
actionable data to conclusively judge the merits of our proposal. 
\end{abstract}



\preprint{ACT-06-12, MIFPA-12-14}

\maketitle


\section{Introduction}

The quest to identify supersymmetry (SUSY) at the Large Hadron Collider (LHC) is advancing
rapidly, where an integrated luminosity of 4.7 \fb at $\sqrt{s} = 7$~TeV was in hand by the
close of 2011, and 2012 offers the prospect of an additional 15 \fb at the richer collision
energy of $\sqrt{s} = 8$~TeV.  The standing prognosis from the ATLAS and CMS Collaborations
is that no excess beyond the Standard Model (SM) expectation has been observed.  In
``{\it Profumo di SUSY}''~\cite{Li:2011av} and ``{\it \ldots Aroma of Stops and Gluinos}''~\cite{Li:2012hm},
we nevertheless argued that a forthcoming discovery of SUSY might already be
actively presaged by the delicate emerging scent of vital excesses in the low statistics collider
data.  Such a distant early warning, if legitimate, would be distinguished from stochastic
``look elsewhere'' styled fluctuations in certain key ways.  In particular, the statistical
significance of observed signal excesses would scale appropriately with upgrades in
luminosity, thus maintaining a consistent window for the extrapolated SUSY mass scale.
Additionally, it would be possible (at least in principle) to produce a succinctly described
physical model that tightly correlates the observed signal strength in those
channels presenting excesses over the expected background while avoiding catastrophic
overproduction in those channels that do not.

In the present study we focus on 42 individual event selection channels from five distinct LHC SUSY searches
conducted by the ATLAS~\cite{ATLAS-CONF-2012-033,ATLAS-CONF-2012-037,ATLAS-CONF-2012-041} and
CMS~\cite{CMS-PAS-SUS-11-020,Chatrchyan:2012qka} collaborations, each belonging to the nominal 5~\fb integrated luminosity
class.  We partition these searches into two categories, based on the value of the signal significance
metric $S/\sqrt{B+1}$.  The significance exceeds a value of 2 for three searches, namely
the ATLAS Hadronic SRC Tight and SRE Loose signal regions of Ref.~\cite{ATLAS-CONF-2012-033}, and
the 7j80 ATLAS Multijet study in Ref.~\cite{ATLAS-CONF-2012-037}.  We undertake a multi-axis $\chi^2$
statistical analysis of these searches within the framework of a model named No-Scale
\fsu5~\cite{Maxin:2011hy,Li:2011rp,Li:2011fu,Li:2011av,Li:2011ab,Li:2012hm}, and compare the resulting
best fit against an earlier study~\cite{Li:2012hm} of the corresponding 1.04-1.34~\fb ATLAS
searches~\cite{Aad:2011qa,Aad:2011ib}.  Remarkably, we find that i) the preferred gaugino mass scale
isolated at the $\chi^2$ minimum has maintained a rock steady correlation across this more than four-fold
increase in data, ii) the individual best fits for each of the high significance searches maintain substantial
mutual coherence, and iii) the statistical preference for No-Scale \fsu5 over the null SM limit has indeed increased
commensurately with the increase in integrated luminosity.  For the remaining 39
searches with $S/\sqrt{B+1} < 2$, we use a parallel $\chi^2$ analysis to establish a lower bound
on the scale of new physics in \fsu5, and demonstrate consistency with the best fit extracted from
those searches possessing a positive data excess.

We do not consider it incidental that the only three 4.7 \fb strategies to transcend a significance
limit of 2 are all of the multijet variety~\cite{ATLAS-CONF-2012-037,ATLAS-CONF-2012-033}. As we have vigorously
suggested for some time~\cite{Maxin:2011hy,Li:2011rp,Li:2011fu,Li:2011av,Li:2011ab,Li:2012hm},
multijet events comprise the most favorable signal region of \fsu5, providing an atypically dominant
SUSY signature that is capable of starkly contrasting the sparse aggregation of expected background contributions.
In spite of the pervasive gloom permeating the high-energy community regarding the prospects for a timely discovery of
SUSY, the lingering perfume of new physics in the 5~\fb class results suggests to us that a critical inflection point
in this debate is rapidly approaching. The persistence or dissipation of this sensation through the 2012 data collection
season will constitute the next vital, and likely conclusive, indication of whether it is the lady herself who draws near.


\section{The No-Scale \bfsu5 Model}

No-Scale \fsu5 (See Refs.~\cite{Maxin:2011hy,Li:2011rp,Li:2011fu,Li:2011av,Li:2011ab,Li:2012hm,Li:2011xu} and
all references therein) is defined by the confluence of the ${\cal F}$-lipped $SU(5)$ grand unified
theory (GUT), two pairs of hypothetical TeV scale vector-like supersymmetric multiplets (flippons) of
mass $M_{\rm V}$ with origins in local ${\cal F}$-theory model building, and the dynamically established
boundary conditions of No-Scale Supergravity. This construction inherits all of the most beneficial
phenomenology of the flipped $SU(5) \times U(1)_{\rm X}$ gauge group structure, as well as all of the
theoretical motivation of No-Scale Supergravity. A more expansive theoretical treatment of \fsu5 can be
found in the cited references, including a comprehensive summary in the Appendix of Ref.~\cite{Maxin:2011hy}.

Supersymmetry must be broken near the TeV scale since mass degenerate superpartners for the known SM
fields are not observed. In the Constrained Minimal
Supersymmetric Standard Model (CMSSM) and minimal supergravities (mSUGRA), this first occurs in a
hidden sector, and the secondary propagation by gravitational interactions into the observable sector
is parameterized by universal SUSY-breaking ``soft terms'' which include the gaugino mass $M_{1/2}$,
scalar mass $M_0$ and the trilinear coupling $A$. The ratio of the low energy Higgs vacuum expectation
values (VEVs) $\tan \beta$,
and the sign of the SUSY-preserving Higgs bilinear mass term $\mu$ remain undetermined, while the
magnitude of the $\mu$ term and its bilinear soft term $B_{\mu}$ are determined by the $Z$-boson mass $M_Z$
and $\tan \beta$ following electroweak symmetry breaking (EWSB). In the most fundamental No-Scale
scenario, $M_0$=A=$B_{\mu}$=0 at the unification boundary, while the entire array of low energy SUSY
breaking soft-terms evolve down with a single non-zero mass parameter $M_{1/2}$. As a result, the particle
spectrum is proportional to $M_{1/2}$ at leading order, rendering the bulk ``internal'' physical
properties invariant under an overall rescaling. The matching condition between the low-energy value of
$B_\mu$ that is required by EWSB and the high-energy $B_\mu = 0$ boundary is extraordinarily difficult to
reconcile under the renormalization group equation (RGE) running. The solution at hand relies on
modifications to the $\beta$-function coefficients that are generated by radiative loops containing the
vector-like flippon multiplets. Via couplings to the Higgs boson, the flippons will also have a direct
impact on the Higgs boson mass $m_h$, producing a 3--4~GeV upward shift that naturally facilitates a
physical Higgs mass in excellent accord with the central ATLAS, CMS, and
CDF/D\O~\cite{Collaboration:2012tx,Collaboration:2012si,TEVNPH:2012ab} signal at 124--126~GeV~\cite{Li:2011ab}.
We emphasize that this Higgs mass range is otherwise rather generically
difficult to reconcile with a light TeV-scale SUSY spectrum.

The \fsu5 model space is adherent to a set of firm ``bare-minimal'' phenomenological
constraints~\cite{Li:2011xu}, including consistency with the world average top-quark mass $m_t$, the
dynamically established boundary conditions of No-Scale supergravity, radiative electroweak
symmetry breaking, the centrally observed WMAP7 CDM relic density~\cite{Komatsu:2010fb}, and
precision LEP constraints on the lightest CP-even Higgs boson $m_{h}$ and other light SUSY chargino and
neutralino mass content. We have further established a highly constrained subspace, dubbed the {\it
Golden Strip}, that conforms to the phenomenological limits on rare processes established by
measurement of the muon anomalous magnetic moment $(g_{\mu}-2)/2$ and the branching ratios of the
flavor-changing neutral current decays $b \to s\gamma$ and $B_S^0 \rightarrow \mu^+\mu^-$. A similarly 
favorable {\it Silver Strip} slightly eases the constraints imposed by $(g_{\mu}-2)$.


\section{No-Scale \bfsu5 Multijets}

The No-Scale \fsu5 model leverages the flippon multiplets to facilitate a flatness in the
running of the $SU(3)$ RGEs ($\beta$-coefficient $b_3$ = 0), blocking
the standard logarithmic enhancement to the color-charged gaugino mass $M_3$
at low energies.  This engenders a distinctive mass texture $M({\widetilde{t_1}}) < M({\widetilde{g}}) < M({\widetilde{q}})$
featuring a light stop and gluino, both substantially lighter than all other squarks. The stability of this
characteristic mass hierarchy is observed across the entire parameter space, a hierarchy that is not, to our knowledge,
replicated in any CMSSM/mSUGRA constructions.  The same strongness of the Higgs to top quark coupling that provides
the primary lifting of the SUSY Higgs mass is utilized in the usual way to generate a hierarchically light
partner stop in the SUSY mass-splitting.

In particular, the light stop $\widetilde{t}_1$ and gluino $\widetilde{g}$ are lighter than the bottom squarks
$\widetilde{b}_1$ and $\widetilde{b}_2$, top squark $\widetilde{t}_2$, and the first and second
generation left and right heavy squarks $\widetilde{q}_R$ and $\widetilde{q}_L$.  \fsu5 thus possesses
a uniquely distinctive test signature at the LHC. This spectrum produces a characteristic event
topology starting with the pair production of heavy first or second generation squarks $\widetilde{q}$
and/or gluinos $\widetilde{g}$ in the initial hard scattering process, with each heavy squark likely to 
yield a quark-gluino pair $\widetilde{q} \rightarrow q \widetilde{g}$ in the cascade decay. The gluino
mediated stops will be off-shell for gaugino masses $M_{1/2} <$ 729 GeV. Specifically, for 600 $< M_{1/2} <$ 729
GeV, the gluino proceeds off-shell via $\widetilde{g} \rightarrow \widetilde{t}_1 \overline{t}$ at a very
high rate of 67-91\%, where the off-shell light stops decay as $\widetilde{t}_1 \rightarrow t \widetilde{\chi}_1^0$
at 52-78\% and as $\widetilde{t}_1 \rightarrow b \widetilde{\chi}_1^{\pm}$ at 13-15\%. On the
contrary, for gaugino masses $M_{1/2} \ge$ 729 GeV, the gluino mediated stops will be on-shell, where here
the gluino decay proceeds on-shell via $\widetilde{g} \rightarrow \widetilde{t}_1 \overline{t}$
at 100\%, where the on-shell light stops decay as $\widetilde{t}_1 \rightarrow t \widetilde{\chi}_1^0$
at 69-76\% and as $\widetilde{t}_1 \rightarrow b \widetilde{\chi}_1^{\pm}$ at 17-22\%.

The impact of these final states is critical. Expectations are that each gluino will produce two to six jets, with the gluino-mediated stop channel producing the maximum of six jets. These processes may then consistently result in a net product of four to twelve jets emergent from a single gluino-gluino pair production event, five to thirteen jets emergent from a squark-gluino pair production event, and six
to fourteen jets emergent from a squark-squark pair production event. After jet fragmentation
following the primary hard scattering events and the sequential cascade decay chain, the final event
distribution will contain an unmistakable signal of high multiplicity jets.


\section{ATLAS 4.7 $\mathbf{{fb}^{-1}}$ Multijet Searches}

The most powerful signal of SUSY will arise where the magnitude of overproduction beyond expectations is
highest. This has been recently exhibited in the search for the Higgs boson, where events are
distributed from 117.5 GeV to 129 GeV for ATLAS and CMS, and 115 GeV to 135 GeV for CDF/D\O. Yet, the
common conviction is that the Higgs boson is more narrowly constrained about 125 GeV, rather than smeared
across some average of the various observed signals, since this specific mass range is where the signal is the
strongest and where the largest number of events compared to the Standard Model have accumulated. In order to likewise focus our attention on the
most potentially active regions of the ongoing SUSY search, we choose to segregate for inclusion in our
best-fit comparative study of No-Scale \fsu5 only those event selection scenarios currently featuring a value
of the signal significance metric $S/\sqrt{B+1}$ larger than 2.

Applying this minimum threshold for signal significance to the full compilation of 5 \fb class studies 
released to date by ATLAS and CMS, only three individual event selections are found to pass: the 7-jet pT$>$80 GeV (7j80)
case of the ATLAS multijets~\cite{ATLAS-CONF-2012-037}, and the SRC Tight and SRE Loose cases of the ATLAS Hadronic
0-lepton search~\cite{ATLAS-CONF-2012-033}. The 7j80 case is notable as the extension to higher luminosity of
an earlier 1.34 \fb ATLAS search strategy~\cite{Aad:2011qa} that was found to be exceptionally favorable in
No-Scale \fsu5~\cite{Li:2011av,Li:2011ab,Li:2012hm}.  Positing an \fsu5 framework, our full expectations
were for 7j80 to likewise be a leading source of excess SUSY event production as
integrated luminosity surged higher. The $S/\sqrt{B+1}$ at 1.34 \fb for 7j80 was measured at
1.12~\cite{Aad:2011qa}. Our expectation for 5 \fb of data for 7j80 was forecast as $S/\sqrt{B+1} = 1.9$ in
Ref.~\cite{Li:2012hm}. The actual 4.7 \fb signal significance computed from the ATLAS observations in
Ref.~\cite{ATLAS-CONF-2012-037} lands at $S/\sqrt{B+1} = 2.07$, representing quite an impressive continuation of
the burgeoning signal first observed in the much more compact data set.

The remaining two 4.7 \fb searches exceeding two standard deviations emerge from ATLAS
Ref.~\cite{ATLAS-CONF-2012-033}, applying 4-jet (SRC Tight) and 6-jet (SRE Loose) cutting
strategies, while concurrently requiring zero leptons. This is intended to isolate the
$\widetilde{g}\widetilde{g}$, $\widetilde{q}\widetilde{g}$, and $\widetilde{q}\widetilde{q}$
channels via $\widetilde{q} \rightarrow q\widetilde{g}$ and $\widetilde{g} \rightarrow q
\overline{q}$. The lepton exclusion here is very important, such that there will typically be no other
productive lepton-free channels capable of generating 4, 5, or 6 jets from the expected pair produced
gluinos and/or heavy squarks.  These studies likewise extend an earlier ATLAS study at lower
luminosity (1.04 \fbns) that had previously attracted our attention.  In this case, we had 
forecast a value of $S/\sqrt{B+1} = 3.0$ in Ref.~\cite{Li:2012hm} for the ``High Mass'' search of
Ref.~\cite{Aad:2011ib}, along with a 24 event excess.  In the prior 1.04 \fb ATLAS 0-lepton analysis of
Ref.~\cite{Aad:2011ib}, the region of large effective mass was essentially consolidated into a single test case,
identified as ``High Mass'', with a cut of $\ge$4 jets and $M_{eff} >$ 1100 GeV. The signal significance of the
``High Mass'' 1.04 \fb search was $S/\sqrt{B+1} = 1.3$. In the 4.7 \fb ATLAS 0-lepton results, ATLAS has
in essence separated the ``High Mass'' search of Ref.~\cite{Aad:2011ib} into the more targeted 4-jet, 5-jet, and 6-jet searches of
Ref.~\cite{ATLAS-CONF-2012-033}. We must therefore analyze the separate 4.7 \fb searches in
Ref.~\cite{ATLAS-CONF-2012-033} to permit a direct comparison to the 1 \fb total $\ge$4 jet ``High
Mass'' result. In this manner, SRC Tight shows $S/\sqrt{B+1} = 3.22$ with an 8.3 event excess, whereas
SRE Loose gives $S/\sqrt{B+1} = 2.65$ with a 29 event excess. The remainder of the
Ref.~\cite{ATLAS-CONF-2012-033} searches are of a very small signal significance, and therefore can be
neglected in this comparison.  Hence, we again find the projections for signal significances and event excesses
to be in very good agreement with the actual data observations of SRC Tight and
SRE Loose.

Taking pause to reflect upon this correlation for a moment, one would not expect random
fluctuations to produce such an interrelation between these two disparate searches. Both strategies
isolate multijet events of 4 jets, 6 jets, and 7 or more jets, the dominant signal region of \fsu5, with
varying cuts on jet $p_T$. Mutual improvement of signal strength from 1 \fb to 4.7 \fb such as these that
correspond to predictions from \fsu5 simulations hints of a possible underlying structure, with the
accidental intersection of random background fluctuations across a more than four-fold increase in data
being highly improbable.  The targeted enhancement in signal significance for those channels in which
a No-Scale \fsu5 signal is expected to be prevalent is moreover highly suggestive.
If indeed the framework of our Universe is described by the \fsu5 model, then such a correlated growth in
signal strength in all productive channels, multijets in particular, would surely be observed.

\begin{figure*}[htbp]
        \centering
        \includegraphics[width=0.8\textwidth]{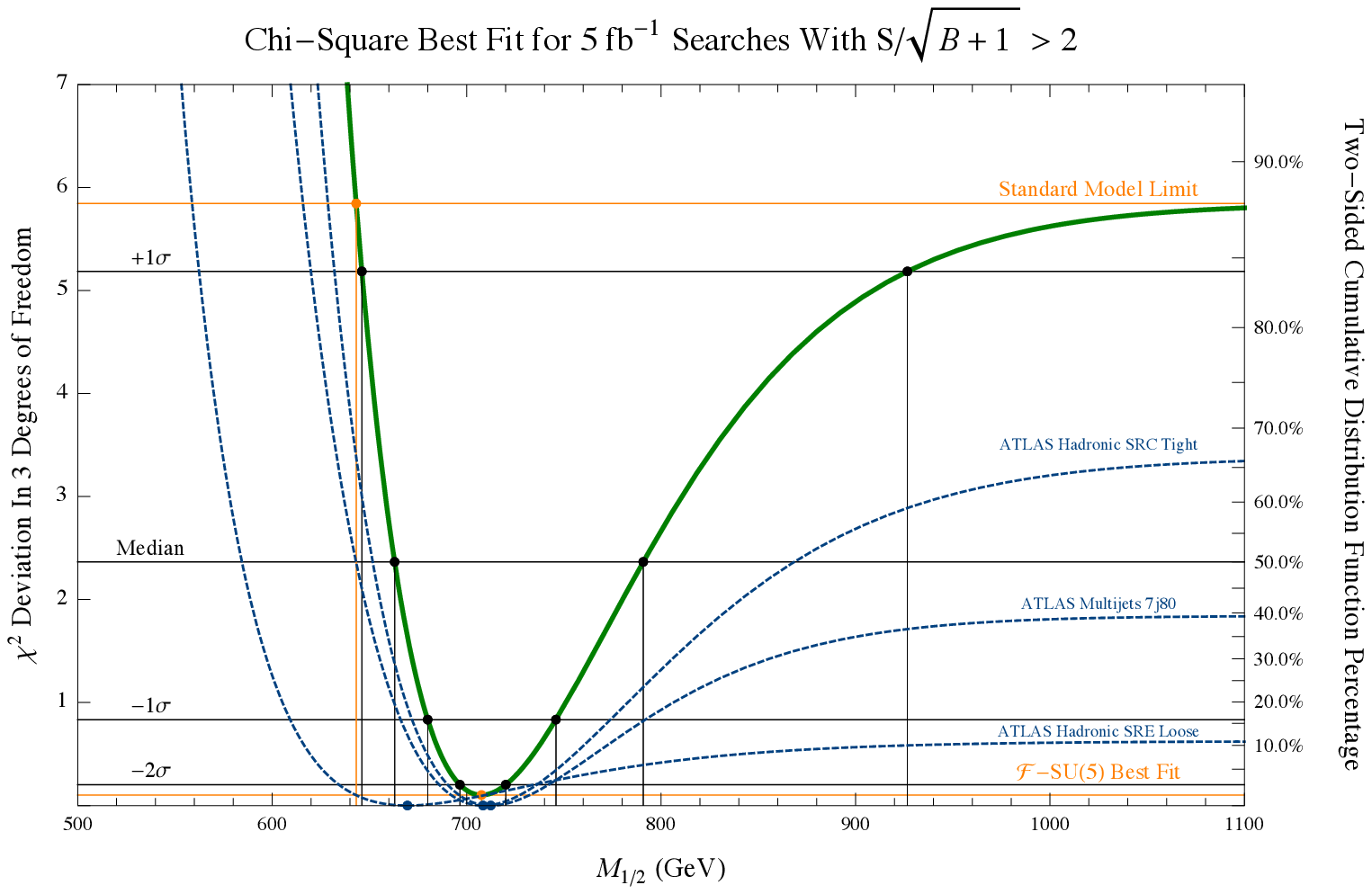} \\
        \includegraphics[width=0.8\textwidth]{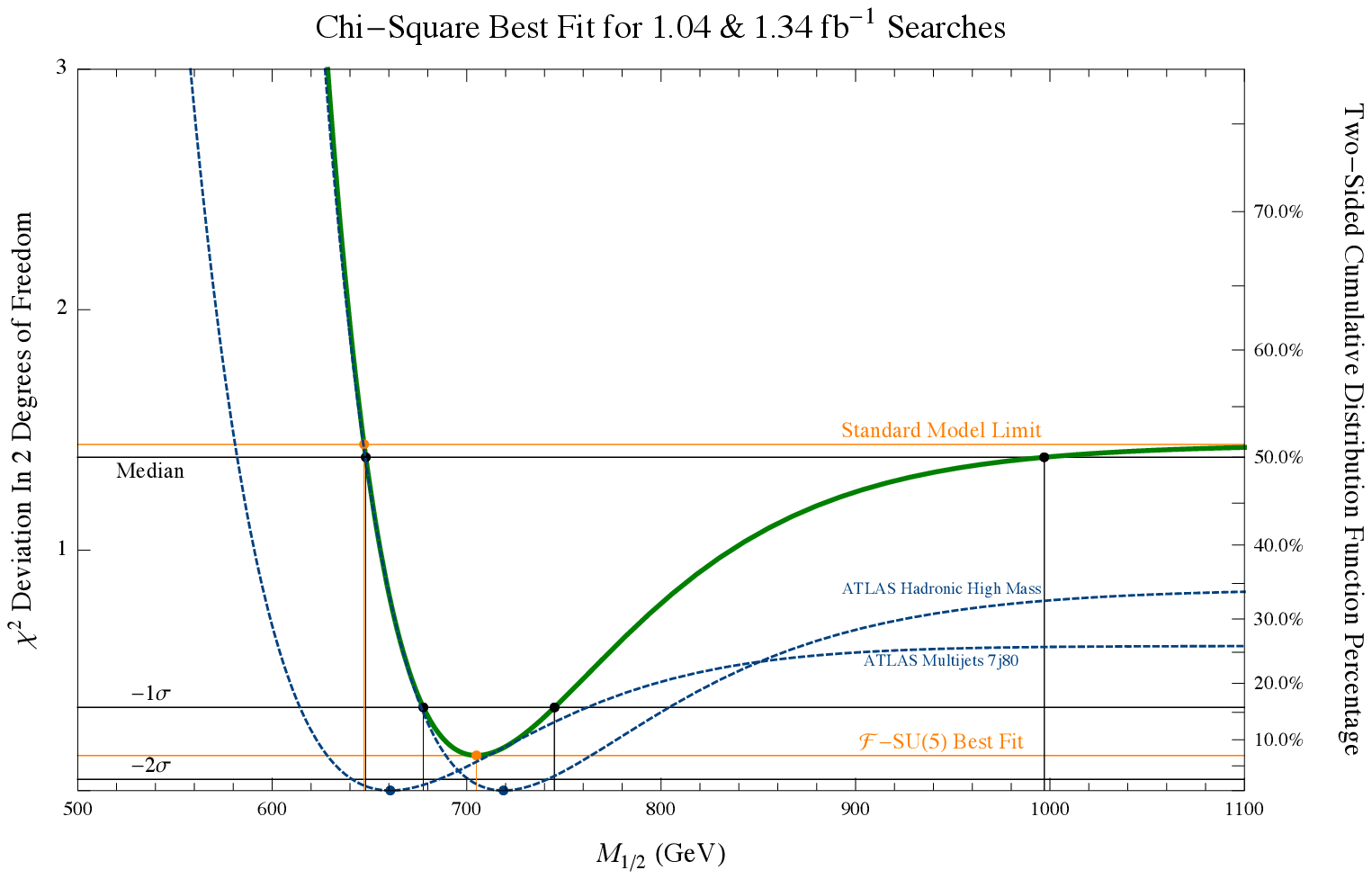}
        \caption{We depict the $\chi^2$ analyses of the ATLAS 4.7 \fb 7j80, SRC Tight, and SRE Loose Multijet search
strategies from Refs.~\cite{ATLAS-CONF-2012-037,ATLAS-CONF-2012-033} (upper pane), and the corresponding historical
studies (``High Mass'' and 7j80 cases of Refs.~\cite{Aad:2011ib,Aad:2011qa}) at 1.04 and 1.34 \fb (lower pane).
The thin dotted blue lines correspond to the individual $\chi^2$ curves for each event selection, which are summed into the
thick green cumulative multi-axis $\chi^2$ curves.  These searches are selected for the exhibition of a signal
significance $S/\sqrt{B+1}$ greater than 2 for the 5 \fb class studies.  A direct visual inspection of the growth of the signal
strength and the fluctuation of the $\chi^2$ minimum with increasing luminosity is thus facilitated.  Remarkably, we
observe extreme stability in the favored mass scale, despite more than a four-fold increase in the integrated luminosity.
The individual curves also display a very tight correlation amongst themselves.  Together with an appropriately large
increase in the ``depth'' of the $\chi^2$ well, we suggest that these features indicate the presence of a non-random structure.}
        \label{fig:CSBF}
\end{figure*}

\section{Chi-Square Analysis}

We apply the $\chi^2$ test statistic to probe for specific correlations between various
ATLAS~\cite{Aad:2011ib,Aad:2011qa,ATLAS-CONF-2012-033,ATLAS-CONF-2012-037,ATLAS-CONF-2012-041} and
CMS~\cite{CMS-PAS-SUS-11-020,Chatrchyan:2012qka} observations and the No-Scale \fsu5 model.
For the error width of each search we combine a statistical factor $\sqrt{S+B+1}$ accounting for
Poisson fluctuations in the net experimental observation in quadrature with the quoted collaboration
estimates of the background uncertainty.  Searches exhibiting an anomalous under-production with
respect to the expected background are zeroed out to allow the full error width for post-SM physics.
Each relevant SUSY search is then compared against the full portion of the \fsu5 model space that
remains simultaneously consistent with all the latest experimental constraints, including a 124-126 GeV Higgs boson
mass, but excluding the ATLAS and CMS SUSY constraints that are the objective of this exercise.
The narrow strip of otherwise viable model parameterizations ranging from 400 $\le M_{1/2} \le$ 900 GeV
is liberally sampled at 22 representative benchmark combinations of $M_{1/2}$, $M_V$, $m_t$ and $\tan\beta$.

For each benchmark sample, we execute an in-depth Monte Carlo collider-detector simulation of all 2-body SUSY
processes based on the {\tt MadGraph}~\cite{Stelzer:1994ta,MGME} program suite, including the {\tt MadEvent}~\cite{Alwall:2007st},
{\tt PYTHIA}~\cite{Sjostrand:2006za} and {\tt PGS4}~\cite{PGS4} chain.
The input SUSY particle masses are computed with {\tt MicrOMEGAs 2.1}~\cite{Belanger:2008sj}, using a proprietary modification of
the {\tt SuSpect 2.34}~\cite{Djouadi:2002ze} codebase to run the flippon-enhanced RGEs.
We use a modified version of the default ATLAS and CMS detector specification cards provided with {\tt PGS4} that
calls on the newly available anti-kt jet clustering algorithm, indicating an angular scale parameter
of $\Delta R = 0.4$ and $\Delta R = 0.5$, respectively.  The resulting event files are filtered according to a precise
replication of the selection cuts specified by the collaborations, employing a script {\tt CutLHCO 2.0} of our own design~\cite{cutlhco}.
Finally, the sampled event counts are used to extrapolate a continuous functional dependence on the gaugino mass $M_{1/2}$
that is suitable for the generation of a $\chi^2$ fitting of the \fsu5 event production against the experimental data.

In order to make a meaningful comparison between our Monte Carlo collider-detector simulation
and the experimental results, it is necessary to establish a consistent cross-calibration.
A convenient language for accomplishing the requisite normalization is
the mutual analysis of a common mSUGRA benchmark.  In most cases, ATLAS and CMS provide
precisely such benchmark data, which is in turn carefully internally calibrated against
their own experimental results.  We find consistent structural correlation against the
major collaboration analyses, affirming the intrinsic soundness of our quantitative procedure.
However, we do observe a small systematic suppression in our absolute event counts relative
to the reference data.  The mean value of the required calibration factor for the four out
of five search methodologies providing a suitable mSUGRA benchmark is 1.84.  When provided dual 
mSUGRA points, we have favored the more conservative adjustment.  The reader
who wishes to compare our current results against prior publications should be aware
that the enhancement in the event production rate garnered by this normalization must
presently be countered by a corresponding suppression, to be achieved via modest elevation
of the mass scale $M_{1/2}$.  In conjunction, we remark again that the No-Scale \fsu5 SUSY
spectrum possesses the rather unique textural characteristic of leading order {\it en masse}
proportionality to just that single dimensionful parameter.  In this sense, the internal physics
of the model are largely invariant under a numerical relabeling of the horizontal $M_{1/2}$ axis.

\begin{figure*}[htb]
        \centering
        \includegraphics[width=0.8\textwidth]{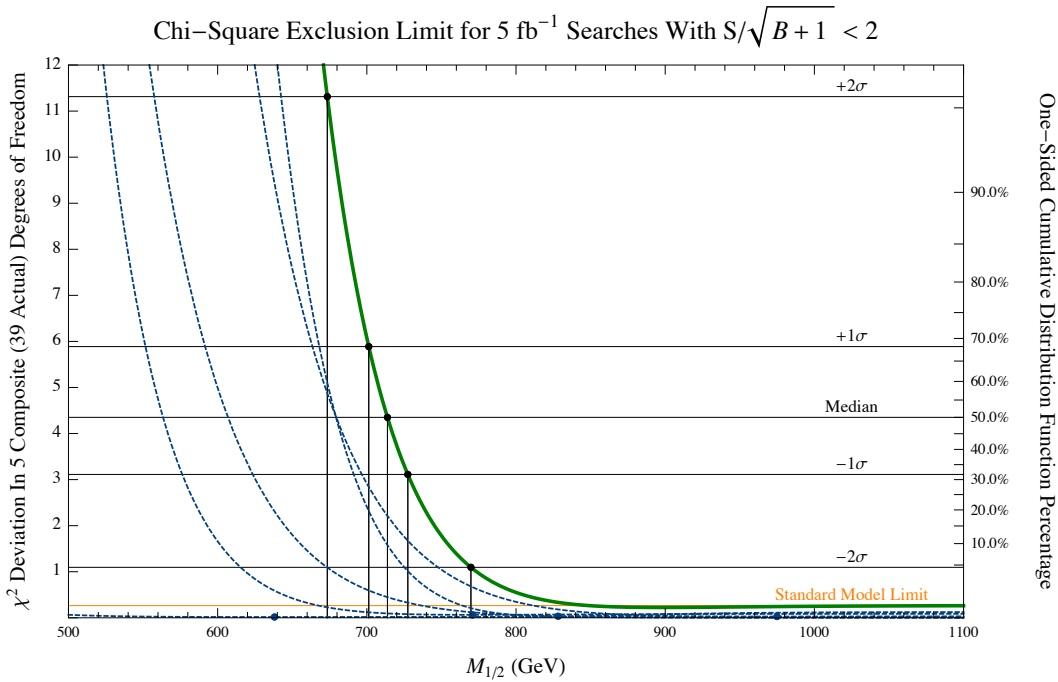}
        \caption{We depict a $\chi^2$ analysis of the 5 \fb class ATLAS and CMS studies
from Refs.~\cite{ATLAS-CONF-2012-033,ATLAS-CONF-2012-037,ATLAS-CONF-2012-041,CMS-PAS-SUS-11-020,Chatrchyan:2012qka}
that exhibit a signal significance $S/\sqrt{B+1}$ less than 2.  The thin dotted blue lines represent composite unit-strength
$\chi^2$ curves for each of the five references, which are summed into the thick green cumulative multi-axis $\chi^2$ curve.
The intention of this study is the establishment of a lower bound on the \fsu5 SUSY mass scale.  At 2$\sigma$ (95\% confidence),
it appears that we may exclude gaugino masses $M_{1/2}$ below 673~GeV.  This is comfortably consistent with the best fit for
$M_{1/2}$ that is established by a parallel $\chi^2$ analysis of those searches exhibiting post-SM physics at a signal significance
greater than 2, as depicted in Figure Set~(\ref{fig:CSBF}).}
        \label{fig:CSEXCL}
\end{figure*}

The chi-square analyses presented in Figure Set~(\ref{fig:CSBF}) are focused on the isolation of a best
fit to the No-Scale \fsu5 mass parameter $M_{1/2}$ against a subset of selection strategies
exhibiting various degrees of activity beyond the SM expectation.
The contemporary searches that we consider to hold data excesses include the 4.7 \fb 7j80 search of
Ref.~\cite{ATLAS-CONF-2012-037} along with the 4.7 \fb SRC Tight and SRE Loose cases of
Ref.~\cite{ATLAS-CONF-2012-033}.  The best fit obtained by this analysis (upper pane) is compared to that
obtained from the corresponding historical (lower pane) 1.04 \fb and 1.34 \fb studies~\cite{Aad:2011qa,Aad:2011ib}.
We will expound upon the remarkable correlation that is exhibited across these dramatically different
luminosities in the next section. Since the best \fsu5 fit in these cases may, in principle, be either better
or worse than the null production scenario embodied in the SM limit, we adopt a two-sided presentation
of the cumulative distribution function.  Each event selection strategy contributing to the
net $\chi^2$ statistic is individually labeled, and individually calibrated against reference
data provided by the ATLAS collaboration.

In the analysis presented in Figure~(\ref{fig:CSEXCL}), we are instead
intent on establishing a global lower bound on $M_{1/2}$ by consideration of those studies without
compelling hints of new physics.  Specifically, we include the 39 signal regions from the 5 \fb class
references under analysis~\cite{ATLAS-CONF-2012-033,ATLAS-CONF-2012-037,ATLAS-CONF-2012-041,CMS-PAS-SUS-11-020,Chatrchyan:2012qka}
that possess a data significance $S/\sqrt{B+1}$ of less than 2.
As such, we adopt a single-sided cumulative distribution function
for this case, and are interested in the values of $M_{1/2}$ at the median, $+1 \sigma$ and $+2\sigma$
intersections of the $\chi^2$ statistic.  Specific correlations exhibited against the best
fit $\chi^2$ analysis are likewise a topic for discussion in the following section.
Expecting that parallel event selections within a single
SUSY search strategy may be strongly inter-dependent, we condense each of the five searches under
consideration into a single unit-strength composite degree of freedom, collectively representing
the complete set of 39 individual event selection channels.  Each composite is calibrated according to a
mean of factors sampled from its constituent channels, when available.  Since no mSUGRA benchmark sample is provided
for the CMS Jets and Dilepton study~\cite{CMS-PAS-SUS-11-020}, the mean of means is applied in this case.

\section{The Correlation of No-Scale \bfsu5 with LHC Observations}

Upon review of Figure Set~(\ref{fig:CSBF}), several features immediately jump out. Firstly, the overall
$\chi^2$ best fit for the 4.7 \fb $S / \sqrt{B+1} \ge 2$ searches at $M_{1/2} = 708$~GeV is in fine accord with
the overall $\chi^2$ best fit at $M_{1/2} = 705$~GeV for the 1 \fb incarnations of the same searches.
Table I gives a benchmark SUSY spectrum for this 4.7 \fb cumulative $\chi^2$ best fit of $M_{1/2}$ = 708 GeV.
Incidentally, the more broadly based $\chi^2$ analysis of Ref.~\cite{Li:2012hm} for seven 1--2 \fb studies has a
calibrated best fit of $M_{1/2} = 689$~GeV, which is likewise in excellent agreement the newer 5 \fb results.
Secondly, there is also steady coherence amongst the individual $\chi^2$ best fits,
with the most productive 4.7 \fb studies (those with $S/\sqrt{B+1} > 2$) in particular exhibiting
no substantial outliers to the global best fit.  Thirdly, the ``depth'' of the $\chi^2$ well is substantially
enhanced in the larger luminosity study.  By way of comparison, the 1 \fb best fit occurs at $\chi^2 = 0.15$ (out
of two degrees of freedom), with a cumulative distribution percentage (CDP) of 7.0\%, while the 4.7 \fb best
fit occurs at $\chi^2=0.10$ (out of three degrees of freedom), with a CDP of 0.8\%.  Similarly, the null SM
limit at 1 \fb occurs at $\chi^2=1.44$ with a CDP of 51.3\%, while the same limit for 4.7 \fb corresponds
to $\chi^2=5.85$ with a CDP of 88.1\%.  It seems that the gulf separating new physics from the SM alternative
is indeed widening with accumulating statistics, exactly as it must if legitimate SUSY production is indeed
at the root of the observed overproduction.  Should this trend continue, it may not be long until the
prospect of a satisfactory SM limit has been methodically laid to rest.

\begin{table}[ht]
  \small
    \centering
    \caption{Spectrum (in GeV) for $M_{1/2} = 708$~ GeV, $M_{V} = 3215$~GeV, $m_{t} = 174.4$~GeV, $\tan \beta$ = 22.22. Here, $\Omega_{\chi}$ = 0.1138 and the lightest neutralino is greater than 99\% Bino.}
		\begin{tabular}{|c|c||c|c||c|c||c|c||c|c||c|c|} \hline
    $\widetilde{\chi}_{1}^{0}$&$143.4$&$\widetilde{\chi}_{1}^{\pm}$&$306$&$\widetilde{e}_{R}$&$264$&$\widetilde{t}_{1}$&$786$&$\widetilde{u}_{R}$&$1371$&$m_{h}$&$124.4$\\ \hline
    $\widetilde{\chi}_{2}^{0}$&$306$&$\widetilde{\chi}_{2}^{\pm}$&$1134$&$\widetilde{e}_{L}$&$741$&$\widetilde{t}_{2}$&$1262$&$\widetilde{u}_{L}$&$1490$&$m_{A,H}$&$1227$\\ \hline
    $\widetilde{\chi}_{3}^{0}$&$1131$&$\widetilde{\nu}_{e/\mu}$&$737$&$\widetilde{\tau}_{1}$&$151$&$\widetilde{b}_{1}$&$1228$&$\widetilde{d}_{R}$&$1421$&$m_{H^{\pm}}$&$1230$\\ \hline
    $\widetilde{\chi}_{4}^{0}$&$1133$&$\widetilde{\nu}_{\tau}$&$718$&$\widetilde{\tau}_{2}$&$725$&$\widetilde{b}_{2}$&$1351$&$\widetilde{d}_{L}$&$1492$&$\widetilde{g}$&$952$\\ \hline
		\end{tabular}
		\label{tab:masses}
\end{table}

Turning attention to Figure~(\ref{fig:CSEXCL}), it is incumbent upon us to verify that the best fit established
in the prior figure is not undone by a fundamental inconsistency with exclusion limits on $M_{1/2}$ from the
studies without any dramatic post-SM production.  At a 2$\sigma$ (95\% confidence) level, it seems that we
may exclude values of $M_{1/2}$ less than 673~GeV.  The 1$\sigma$ and median intersections with the $\chi^2$
curve occur at 701 and 714~GeV, respectively.  This appears to provide a quite satisfactory overlap with
the intersection boundaries of the median fit for the upper pane of Figure Set~(\ref{fig:CSBF}) at 663--790~GeV.
Of course, indefinitely large values of $M_{1/2}$ are no worse a fit to the data represented by this figure
than the SM limit itself.

We cannot help but entertain optimism for an imminent SUSY discovery in the tranche of data to be
released in 2012. With another 15 \fb at $\sqrt{s}$ = 8 TeV expected to be delivered over the running season,
the debate of whether we live in a supersymmetric universe could reach a climactic resolution soon. Assuming the next
release of LHC data in Summer 2012 is 5 \fb at 8 TeV, for an interim total of 10 ${\rm fb^{-1}}$, a quick ``back of the envelope''
projection suggest that discovery might already be within sight, even at that intermediate point.
Specifically, our simulations of the key signal space at 8~TeV suggest that the increased beam energy
may yield a multiplicative signal efficiency advantage of around 3.66.  Na\"ively extrapolating this same
advantage onto the 6.4 events observed over a background of 8.6 events in the ATLAS 7j80 search of
Ref.~\cite{ATLAS-CONF-2012-037}, we might reasonably expect a signal significance for this channel
on the order of $4.66\times6.4/\sqrt{4.66\times8.6+1} \sim 4.7$ with 5 \fb each of 7~TeV and 8~TeV data. Likewise, for the 0-lepton search, we find a multiplicative signal advantage of about 3.1 for 10 \fbns, yielding an $S/\sqrt{B+1}$ signal significance near 7.0 for SRC-Tight and 5.4 for SRE-Loose.
In light of this, we eagerly await the first 8 TeV data release by the ATLAS collaboration, and particularly
so for these sets of event selection cuts.

In closing, we wish to briefly highlight the 4.7 \fb cumulative $\chi^2$ best fit lightest supersymmetric particle (LSP) mass of
$m_{\widetilde{\chi}_1^0}$ = 143.4 GeV. While the greater than 99\% bino composition of the \fsu5 LSP
generates a much smaller photon-photon annihilation cross-section and gamma-ray flux (when using the
typical Einasto halo profile assumptions) than the FERMI-LAT reported fluxes of
Refs.~\cite{Bringmann:2012vr,Weniger:2012tx}, we do nonetheless find the tentative measurement of a
130 GeV monochromatic gamma-ray line at over 4$\sigma$ quite interesting. Adding
even further intrigue is the more recent result~\cite{Tempel:2012ey} that due to energy loss from final
state radiation in the $\gamma \gamma$ final state, the $\chi^2$ best fit to the 130 GeV monochromatic gamma-ray line is in fact $M_{DM} = 145$ GeV.


\section{Conclusions}

The overarching message that the reader should take from the analysis presented in this work is the existence of a beautiful
correlation between the ATLAS 1 \fb to 4.7 \fb multijet observations, and the intrinsic consistency with which certain
event selection channels are currently exhibiting overproduction at the LHC. We partitioned the 5 \fb class searches
into two categories based on the signal significance metric $S/\sqrt{B+1}$.  Those searches with a value greater than 2,
all corresponding to ATLAS Multijet studies (4 jets, 6 jets, and $\ge$7 jets), were used to establish a $\chi^2$
best fit to the SUSY mass scale in the context of the No-Scale \fsu5 model.  The revealing aspect of these three searches is that they all
reside in the heart of the \fsu5 signal region, which lays claim to multijets as an atypically dominant
signature.  We discovered that in the realm of an \fsu5 framework, a splendid correlation exists between the best fits of
the 1 \fb and 4.7 \fb $\chi^2$ studies, and all high significance searches moreover
possess an individual best fit in close proximity to the overall best fit. Such intricate natural
correlations hint of an underlying structure, and not unpredictable randomly delivered fluctuations of
the background.  Our proposed benchmark has a Higgs boson mass $m_{h}$ = 124.4 GeV, and a best fit SUSY spectrum
featuring an LSP mass $m_{\widetilde{\chi}_1^0}$ = 143.4 GeV, light stop mass $m_{\widetilde{t}_1}$ = 786 GeV, gluino mass $m_{\widetilde{g}}$
= 952 GeV, and $u_L$ heavy squark mass $m_{\widetilde{u}_L}$ = 1490 GeV.

The grand finale of this longstanding debate over the reality of supersymmetry in our Universe could
arrive rather soon, materializing well before most high-energy physicists imagined, particularly
pessimistic critics hardened by the early demise of the most popular SUSY constructions. Notably, we suggest
that if current signal strengths hold up, the next release of LHC data, assumed to be 5 \fb at 8 TeV, could
yield signal significances very close to 5 for certain individual highly favored production channels.
Such an eventuality would dramatically expose the already plausible possibility that supersymmetry is
in fact quite alive and well at the LHC, robustly amassing statistics as the integrated luminosity surges
steadily upward. As we await this ever important next assemblage of data, we again
accentuate the point that the validation of an \fsu5 framework underlying the LHC collisions would necessarily also carry
with it implications more profound than even just the realization of supersymmetry, touching also on the stringy origins of
our Universe, the landscape of string vacua, and even perhaps the No-Scale foundations of the Multiverse.
But those are stories for another day; the day after the conclusive discovery of supersymmetry.


\begin{acknowledgments}
This research was supported in part
by the DOE grant DE-FG03-95-Er-40917 (TL and DVN),
by the Natural Science Foundation of China
under grant numbers 10821504, 11075194, and 11135003 (TL),
by the Mitchell-Heep Chair in High Energy Physics (JAM),
and by the Sam Houston State University
2011 Enhancement Research Grant program (JWW).
We also thank Sam Houston State University
for providing high performance computing resources.
\end{acknowledgments}


\bibliography{bibliography}

\end{document}